\documentclass[runningheads]{llncs}
\usepackage{graphicx}
\RequirePackage{amsmath,amsfonts,amssymb}
\usepackage{multirow}
\usepackage{xcolor}
\begin{document}
\title{Modeling Memory Imprints Induced by Interactions in Social Networks}
%
%
\author{James Flamino\inst{1} \and
Ross DeVito\inst{2} \and
Omar Lizardo\inst{3} \and
Boleslaw K. Szymanski\inst{1}}
\authorrunning{J. Flamino, R. DeVito, O. Lizardo, B.K. Szymanski}
%
\institute{Rensselaer Polytechnic Institute, Troy NY 12180, USA \and
University of California, San Diego, CA 92093, USA \and
University of California, Los Angeles, CA 90095, USA \\
$^{\dagger}$ This paper is a preprint version of a publication of the same title presented at the SBP-BRiMS 2022 conference. Wording may vary between versions.
}
\maketitle              
\begin{abstract}
Memory imprints of the significance of relationships are constantly evolving. They are boosted by social interactions among people involved in relationships, and decay between such events, causing the relationships to change. Despite the importance of the evolution of relationships in social networks, there is little work exploring how interactions over extended periods correlate with people's memory imprints of relationship importance. In this paper, we represent memory dynamics by adapting a well-known cognitive science model. Using two unique longitudinal datasets, we fit the model's parameters to maximize agreement of the memory imprints of relationship strengths of a node predicted from call detail records with the ground-truth list of relationships of this node ordered by their strength. We find that this model, trained on one population, predicts not only on this population but also on a different one, suggesting the universality of memory imprints of social interactions among unrelated individuals. This paper lays the foundation for studying the modeling of social interactions as memory imprints, and its potential use as an unobtrusive tool to early detection of individuals with memory malfunctions.

\end{abstract}
\section{Introduction}

Interactions among people in social networks drive their temporal evolution. Yet, the underlying mechanics of mapping interaction patterns to the evolution of relationships, and consequently to personal memory imprints, are not clear. A primary reason is the paucity of data providing comprehensive information on both interactions (such as Call Detail Records) and relationship characteristics (such as ego network surveys). Previous work using such data focuses instead on machine learning and natural language processing models, or more ``black box'' approaches ~\cite{gilbert2009predicting,conti2011model,flamino2021machine}. While this work may accurately predict surface-level correlations among interactions and people's memory imprints, they lack any analytical, explanatory value. An individual’s memory recall reflects their perception of the impact of interactions, each leaving a temporally evolving imprint. But how can such longitudinal phenomena in social networks be directly modeled in analytical form? 

In a recent paper~\cite{michalski2021social}, Michalski et al. implement a cognitive model of memory recall (CogSNet) and evaluate its accuracy at ``nominating'' significant relationships between individuals from dyadic communication records. This research provides insight into how patterns of communication map to strengths of memory recall of interacting members of the network. However, the authors formulate their model with a parameter which is not a part of the established, well-known memory model,  ACT-R~\cite{anderson1998integrated}. This added parameter represents the threshold value of a memory imprint at which such imprint vanishes from memory. Yet, unexpectedly, optimal values of this parameter are high and are unlikely to correspond to human memory processes. In this paper, we modify CogSNet, improving its compatibility with ACT-R. CogSNet uses three parameters to represent memory imprints. Our adapted model, the Memory Imprint Model (MIM), inherits two CogSNet parameters based on the ACT-R model that is applied to the entire population. We replace the third CogSnet parameter with a new one that defines, for each individual, a threshold for the significance of relationships with their partners. This change is significant since we discovered that the optimal fit for the MIM parameters are actually universal for populations of peoples with healthy memories. Provided that individuals with malfunctioning memory will recall past interactions and relationships differently than healthy ones, we hypothesize that MIM could be an unobtrusive tool for diagnosing memory malfunction in humans. 

In the rest of the paper, section \ref{sec:data} outlines the datasets used to evaluate MIM. Then, in section \ref{sec:models}, we explain how MIM differs from CogSNet, and how we map interactions among individuals into an evolving memory imprint. We compare the performance of MIM to a standard Hawkes model and three baseline models. Section \ref{sec:res} discusses our evaluation process in detail. In section \ref{sec:cross}, we assess the universality of model parameters and find that parameters learned on a population of healthy subjects are valid for other populations of similar individuals. Finally, section \ref{sec:conc} summarizes all our results and hypotheses, discusses their implications, and proposes some avenues for future research.

\section{Datasets}
\label{sec:data}
Testing the mechanisms of memory imprints arising from interactions in social networks requires longitudinal data containing objective records of communications among individuals within a social network, as well subjective data on relationship characteristics, such as those obtained using ego network surveys. The former is used to provide dyadic events that can be fitted by the model to simulate the memory imprint left by interactions between the two involved individuals. The latter is used to provide actual empirical measures of significance between the two people involved as they see it. Unfortunately, data fitting these requirements is scarce. Of the past works mentioned above, most of those having data meeting these requirements are cross-sectional, thus only capable of testing their models on snapshots of social networks but unable to explore the evolution of those networks. One exception is the already mentioned CogSNet model that uses data collected in the NetSense study~\cite{striegel2013lessons}. The data were voluntarily collected from randomly selected first-year students entering the University of Notre Dame in 2011. From Fall 2011 to Spring 2013, the participating students had Call Detail Records of their phones continuously recorded. In addition, the participants were required to fill out ego network surveys at the end of each semester. The NetSense study followed 196 students at its peak. We also use data from the follow-up NetHealth study~\cite{purta2016experiences} involving first-year students arriving in Notre Dame in 2015 and leaving in 2019 using a similar approach as used by NetSense. NetHealth followed 594 students at its peak.

\subsection{Communication Records} 

Communication records for both studies were formatted to conform to the standard CDR format that includes for each call a timestamp, sender, receiver, type, and length. While there are auxiliary digital communication channels in NetSense (such as WhatsApp), these are sparse, so we keep only the text and phone calls, which provided us with $7,465,776$ events. We filter similarly for NetHealth, extracting $41,677,368$ calls and text messages generated by the participants in that study.

\subsection{Ego Network Surveys} 

Each study contains ego network surveys collected once a semester to complement communication records. Ego network surveys in these studies essentially capture student perceptions on their interactions, which characterizes their subjective rapport with those involved. These surveys were prefaced with a question (the name generator) asking the survey-taker (the ego) to list individuals (the alters) with whom they spend a significant amount of time communicating or interacting. Each ego can list up to 20 alters, including people not participating in the study. Thus, ego-alter connections can involve a variety of relationship types such as friends, parents, coworkers, and romantic partners. Each ego was subsequently asked about the history of contact with these alters and their shared interests and activities (the name interpreter). Importantly, the surveys also asked the ego to subjectively rate similarity and closeness with the alters. The NetSense study includes four ego network surveys over four semesters, and the study participants listed on average $13.9$ alters per survey. The NetHealth study includes eight ego network surveys over eight semesters, and the study participants listed on average $12.06$ alters per survey.

\subsection{Quantifying Relationship Significance} 

To model the significance of interactions using memory imprinting, we need an empirically-based measure of the significance of relationships that acts as ground truth. Previously, in~\cite{michalski2021social}, the authors establish their ground truth as the list of alters nominated by the ego. According to the authors: ``These nominations are based on students' perception of the corresponding relations as one of the top twenty most interacting peers in the surveys administered to participants.'' However, while the set of the alters themselves can be considered significant to the ego as a whole, there is additional information from the surveys that provide information on the significance of individual alters, as perceived by the egos, in even greater detail. This allows us to produce a finer measure of significance from which we can more thoroughly test the capabilities of the models in this paper.

In the previous section, we mentioned that there are ``name interpreter'' questions asked about the alters that are listed by the ego. The answers to these questions are mostly selected from a set of modalities that grade the magnitude of an attribute of the alter or the relationship. For example, when asking about an ego's perceived closeness with a listed alter, the choices include ``Especially close'', ``Close'', ``Less than close'', or ``Distant''. There are also a few questions requiring open answers in the form of a rational number (like self-reported tie duration). In this paper, we focus on a subset of the inputs from the surveys: closeness, history duration, subjective similarity, emotional significance (for the NetSense surveys only), and perceived communication frequency (for the NetHealth surveys only). With these question sets, we convert the list of alters into a ranked list, where the higher the rank of an alter, the more significant they are to the ego. 

To determine an ordering from the mixed inputs without assuming the importance of any one input over another, we use a pairwise comparative tournament selection process. Consider one ego network survey. Every alter listed by this ego is compared on each question against all the other listed alters. An alter whose assigned answer to a question is considered ``closer to the ego'' compared to their counterpart, is awarded a point. If the answers of both competitors are of the same value, they are both awarded a point. These points are aggregated across all the questions and then a ranked list is created by ordering everyone by their score in descending order. If there is a tie in the aggregate score between two alters, the duration attribute is used to break the tie, as it is the only considered input with a rational number answer. After all comparisons are complete for that ego network survey, we are left with our rankings for the top-$k$ significant dyadic relations between the ego and each of its listed alters. The higher the alter is on the list, the more significant relations this alter has with the ego. We run this process for all egos for every survey, yielding our ground truth for this paper.

Given a model and CDR data between an ego and an individual, who interacted with this ego, we fit our model to predict not only if this individual should be listed on the ego survey list, but also what rank, if any, this individual should have on this list. This allows us to assess the model's ability to determine finer levels of significance from the memory imprint from events, not just if the considered relationship has perceived significance.

\section{Models}
\label{sec:models}

\subsection{Hawkes Model}

The Hawkes process is a natural choice for simulating the significant relations between two individuals as determined by communication, since the frequency and volume of communication is central to determining the significance of the relationship. Furthermore, past work directly uses this method to understand relationship dynamics~\cite{nurek2020hawkes}. Here, we use a univariate Hawkes process with an exponential kernel for alter nomination and ranking, as defined by the Eq.~\ref{eq:hawkes_comb}.

\begin{equation}\label{eq:hawkes_comb}
    \lambda(t) = \lambda_{0}(t) + \sum\limits_{t_i < t} \beta e^{-\beta (t - t_i)}
\end{equation}

where $\lambda$ is the Hawkes event intensity at time $t$ (the event time at which the function is being evaluated), $t_i$ is a set of event times before the event time $t$, and $\beta$ is the decay rate (which in this scenario represents the decay rate of the memory imprint of an event). We set the arrival rate of immigrant events $\lambda_{0}(t)$ to $0$. Given this function, at survey time $t$, we consider the communication history of all individuals who have communicated with the ego (as at this point we do not know who is an alter). We feed this data into the Hawkes process up to time $t$, producing a corresponding $\lambda$ value for each candidate alter. We then sort these candidate alters in order of descending magnitude of their associated $\lambda$ value, producing an ordered list of who is significant to the ego according to Hawkes event intensity. We discuss tuning the Hawkes model parameters in subsection~\ref{sec:evalp}.

\subsection{Memory Imprint Model}

Our next model, the Memory Imprint Model (MIM), is a variant of the CogSNet model introduced in~\cite{michalski2021social}. Unlike Hawkes, CogSNet uses a recursive piecewise function (cf. Eq.~\ref{eq:cogsnet}) and produces a signal that is normalized to the range $[0,1]$. We made several adjustments to CogSNet to improve its compatibility with the ACT-R model, and to reduce its computational complexity. First, we define the MIM recall function:

\begin{equation} \label{eq:recall}
    r_{ij}(t) = w_{ij}(t^b_{ij})e^{-(t - t^b_{ij})\frac{\ln{(2)}}{L}}
\end{equation}

where $t^{b}_{i,j}$ is the time of the last communication event between nodes $i$ and $j$ in the considered social network before the evaluation time $t$, and $L$ is the parameter that controls the half-life of node $i$'s memory imprint of their interactions with $j$. Given this, we define the strength of the MIM memory imprints of interactions between $i$ and $j$ to be

\begin{equation} \label{eq:cogsnet}
    w_{ij}(t) = s_{ij}(t) + 
        \begin{cases}
        r_{ij}(t)(1 - s_{ij}(t)) & \text{when $t > t_{ij}$} \\
        \mu + (1-\mu)r_{ij}(t)(1 - s_{ij}(t)) & \text{otherwise}
    \end{cases}
\end{equation}

where $\mu$ is the MIM second parameter that defines the memory imprint strength between $i$ and $j$. We list alters $j$'s of ego $i$ at each survey time $t$ in the descending order of their $w_{i,j}(t)$. The length of the list is limited to 20, to mimic the limit for the list length in surveys, and by the threshold defined by the third MIM parameter $\theta$, which limits eligible for listing alters to those whose value of $w_{i,j}(t)$ exceeds $\theta$. 
Ultimately, MIM introduced here is simpler than CogSNet, and fully compliant with ACT-R. We discuss tuning the MIM parameters in subsection \ref{sec:evalp}.

\subsection{Baseline Models}

Previous research has often used frequency of communication to predict significance of relations~\cite{conti2011model,gilbert2009predicting,bulut2012exploiting,flamino2021machine}. Following this established approach, we specify the frequency model in the baseline class that calculates frequency (this model's value that is used to nominate significant alters and rank them) by dividing the number of communication events between two individuals by the elapsed time since they first communicated at the timestep for which it is being evaluated. In addition, we create a recency model using the elapsed time since the last communication between two people as an inversely related estimate of frequency of contact, as is done in~\cite{conti2011model}. This produces a recency value that we use to nominate significant relationships and generate the corresponding rankings. Finally, we create a random model that just randomly orders all individuals that the target ego has communicated with, prior to the considered time, into an arbitrary ranking.

\section{Results}
\label{sec:res}

\subsection{Evaluation Process}
\label{sec:evalp}
Given our two datasets are longitudinal, each with its own set of egos, list of alters, and ego-alter CDR, we designed a comprehensive evaluation process that provides training data for the models to fit their parameters and testing data for the models to test their fit. In the process, we avoid any overlap between training and testing data or accessing in learning the future CDR or ground truth data. 

Initially, for a given dataset, we use the standard three-fold cross validation to shuffle and then equally split the involved egos into three mutually exclusive groups. Then, for each fold, let $N_s$ be the number of semesters within the target dataset. Within a fold, we separate the training and testing data into $N_s$ subsets, split by semesterly survey time. At each survey time, the training and testing subset includes only the surveys and communication data from before that time (therefore the later subsets contain the earlier subsets). We stagger the communication training data, such that the models only fit their parameters with communication data up to the survey time that precedes the survey time against which the models are being tested. This design ensures that the models cannot use any of the training surveys administered at the same time as the current target testing surveys. Subsequently, we start with the second semester for rankings in each fold. After models have fit their parameters with the current training subset and have produced rankings for the current testing subset, the process is then advanced one subset forward, allowing more training and testing data to be released with each following survey time.


Note that the training and testing data are kept completely separate at all times. Testing data, even from previous semesters, is never used to fit the models' parameters, and test-fold ground-truth from surveys is only used to score a model's rankings. The order of the training and testing data used in each fold is kept the same for all models to ensure training and testing consistency. Once all predictions for a fold are complete, we compute the fold's overall score: the weighted average ranking prediction accuracy of the predicted rankings across all available surveys for testing egos. 

For ranking accuracy when comparing rankings generated by a model against the top-$k$ ground truth, we use Ranked Biased Overlap (RBO)~\cite{webber2010similarity}. We chose it because of its use of an indefinite rank similarity measure for the similarity of two ranked lists. Moreover, RBO handles items being present only in one ranking, weights the agreement of top ranking items higher than that of lower ranking items, and works with any given ranking length. This makes RBO a better measure of accuracy for our models' rankings than Jaccard Similarity or Kendall's $\tau$. We set the RBO's weight parameter $p$ to $0.98$. 

The average of the RBO scores for all available testing ego surveys is weighted by the length of the corresponding top-$k$ ground truth lists, accentuating accurate predictions of larger lists. The final score is computed as the average fold score over all folds. This setup allows our models to show if they can continuously predict rankings over extended periods.

When fitting the models using the training data, we use a proven, established framework for optimized parameter fitting: Optuna~\cite{optuna_2019}, which uses an efficient approach to parameter sweeping and critical parameter identification when tuning models. Optuna is used to tune $\beta$ for Hawkes, and $L$, $\mu$, and $\theta$ for MIM each time a new chunk of training data is released during the evaluation process.

\subsection{Model Performance}

We evaluate the performance of each model on NetSense and NetHealth data and present the results in Table~\ref{tab:performance}. The random model sets the lowest bar for performance, followed by recency and the frequency. The tunable models, Hawkes and MIM, outperform the baselines to a statistically significant degree but these two are statistically equivalent. Both models are promising, considering the fact that these models are not only nominating significant relationships for the fit of their interactions, but are also assigning accurate levels of significance of each relationship.

\begin{table}
\centering
\caption{Average RBO scores (variances) for NetSense and NetHealth.}
\label{tab:performance}
\begin{tabular}{|l|l|l|} 
\hline
\multicolumn{1}{|c|}{\multirow{2}{*}{\textbf{Model}}} & \multicolumn{2}{c|}{\textbf{RBO Scores}} \\
\multicolumn{1}{|c|}{} & \multicolumn{1}{c|}{\textbf{NetSense}} & \multicolumn{1}{c|}{\textbf{NetHealth}} \\ 
\hline
Random & 0.03022 (0.002) & 0.03299 (0.004) \\
Recency & 0.28295 (0.029) & 0.27508 (0.033) \\
Frequency & 0.29526 (0.030) & 0.29526 (0.030) \\
MIM & 0.36058 (0.031) & 0.35833 (0.036) \\
Hawkes & 0.36497 (0.031) & 0.37117 (0.038) \\
\hline
\end{tabular}
\end{table}

\section{Cross-Dataset Modeling}
\label{sec:cross}
In this section, we evaluate the generalizability of our models by using Hawkes and MIM tuned on data from one study to nominate and rank alters in the other study. Successful cross-dataset ranking performance serves as an indicator that a model is picking up on transferable cognitive mechanisms that should be universal across populations. In this comparison, we need to consider the different sizes of NetSense and NetHealth datasets. Tuning models on NetHealth to test on NetSense will give them an overt advantage over the time series models because of the size of the training data compared to the testing data. As mentioned in section \ref{sec:data}, NetHealth has 594 egos, while NetSense only has 196. Therefore, we randomly remove 6 egos from the NetHealth dataset and then evenly split the remaining egos into three subsets, resulting in three subgroups of 196 egos, equal to the number of available egos in NetSense. Then to equalize the number of social tie rankings, we repeat the splitting twice, with the egos first keeping their first four semesters of data, and then the last four semesters of data. This results in six NetHealth subgroups that can be used independently for training and testing with NetSense. 

Here, we evaluate only the memory models, since they have parameters that can be tuned, and since the two datasets were collected during entirely different times, we do not need to worry about staggering the training and testing process as done previously. When training on NetHealth to test on NetSense, we tune the models on one of the six NetHealth subgroups and then test it against all the NetSense data, then repeat for all subgroups. When training on NetSense to test on NetHealth, we tune the models just once on NetSense and then test on each NetHealth subgroup. For both approaches, the average of all RBOs weighted by top-$k$ list length gives us the subgroup scores, the average of which creates the final score. The results of this evaluation process are shown in Table~\ref{tab:cross_domain_performance}.

\begin{table}
\centering
\caption{The dataset on the left of the arrow is the dataset on which the models were exclusively trained. The dataset on the right of the arrow is the dataset the models were tested on after training. The score is average RBO with the variance in parentheses.}
\label{tab:cross_domain_performance}
\begin{tabular}{|l|r|r|} 
\hline
\multicolumn{1}{|c|}{\multirow{2}{*}{\textbf{Model}}} & \multicolumn{2}{c|}{\textbf{Training Dataset $\rightarrow$ Testing Dataset}} \\
\multicolumn{1}{|c|}{} & \multicolumn{1}{l|}{\textbf{NetHealth $\rightarrow$ NetSense}} & \multicolumn{1}{l|}{\textbf{NetSense $\rightarrow$ NetHealth}} \\ 
\hline
MIM & 0.35970 (0.0005) & 0.33063 (0.0021) \\
Hawkes & 0.36293 (0.0006) & 0.35071 (0.0016) \\
\hline
\end{tabular}
\end{table}

Importantly, the cross-dataset modeling yields results similar to those obtained on within-dataset. This suggests that these tuned models are picking up on memory imprint mechanisms that are universal across datasets. The universality of these mechanisms implies that while communication patterns and overall behavior within populations may change, the way individuals process and perceive them in memory over time does not. Our models are trained on subjects with healthy memories. Hence, we can hypothesize that they are effective not only at modeling memory imprints within single social networks, but also on social networks with healthy members. 

\section{Conclusions}
\label{sec:conc}
Our results show that the Memory Imprint Model and the Hawkes model are capable of effectively capturing memory imprints of interactions. This process not only allows us to identify significant relationships from interactions over time, but also allows us to assign each relationship a level of significance compared to others, giving us a finer degree of explanatory power when evaluating connections among people in social networks. Importantly, our results on the cross-dataset modeling shows that these models, when fit on the memory imprint dynamics of one dataset, can be used to accurately nominate and rank significant relationships of the other dataset. Based on these results, we hypothesize the MIM and Hawkes models capture the dynamics of a healthy memory, as we are studying the cognitive dynamics of populations of young adults starting college. Consequently, the participants whose mechanics of memory imprint we are modeling will be, generally, in their prime health-wise. Hence, the accuracy of the results for cross-dataset modeling confirms the two models' efficacy in representing the mechanics for the memories of young, healthy adults. 

In future work, we will attempt to more directly evaluate the hypothesis that the models and the process we have designed for evaluating said models can be considered a viable standard for representing memory imprint of interactions for healthy memories that can ultimately be used as a comparative tool for diagnosing malfunctioning memory imprint patterns in humans. We believe that perception of healthy and malfunctioning memories differ. Unlike current indirect methods for diagnosing memory ailments~\cite{bayat2021gps}, modeling perception similar to what was done with student perception in this paper allows us to explicitly measure the strength of distortion of perception. Specifically, if we tune our models to nominate significant relationships using a population of healthy adults and then test it against a population where some individuals have malfunctioning memory (e.g., Alzheimer's, Dementia, Mild Cognitive Impairment), the model performance will be negatively affected, as the memory imprints of the ailing individuals would not match the memory imprints of the healthy population. The easily and unobtrusively obtainable differences in memory import distortion may indicate the type and level of advancements of illnesses, like Alzheimer's or dementia.



%
%

\bibliographystyle{splncs04}
\bibliography{bibliography}

\subsection*{Acknowledgments}
JF and BKS were partially supported by the Army Research Office (ARO) under Grant W911NF-16-1-0524 and by DARPA under Agreements W911NF-17-C-0099 and HR001121C0165. Data collection for the NetHealth project was supported by National Institutes of Health grant \#1 R01 HL117757-01A1 (OL Investigator). Data collection for the NetSense project was supported by National Science Foundation grant \#0968529 (OL Co-PI).

\subsection*{Disclaimer}
We acknowledge that this is a preprint of now published work. This preprint has not undergone peer review or any post-submission improvements or corrections. The Version of Record of this contribution is published in ``Social, Cultural, and Behavioral Modeling''~\cite{flamino2021modeling}, conference proceedings for SBP-BRiMS 2022, and is available online at \url{https://doi.org/10.1007/978-3-031-17114-7\_18}.

\end{document}